\documentclass[prd,onecolumn,nofootinbib,showpacs,showkeys]{revtex4}
\newcommand\gothfamily{\usefont{U}{ygoth}{m}{n}}
\DeclareTextFontCommand{\textgoth}{\gothfamily}

\begin{document}

\title{THE MAXWELL LAGRANGIAN IN PURELY AFFINE GRAVITY}

\author{{\bf Nikodem J. Pop\l awski}}

\affiliation{Department of Physics, Indiana University, Swain Hall West, 727 East Third Street, Bloomington, IN 47405, USA}
\email{nipoplaw@indiana.edu}

\noindent
{\em International Journal of Modern Physics A}\\
Vol. {\bf 23}, Nos. 3 \& 4 (2008) 567--579\\
\copyright\,World Scientific Publishing Co.
\vspace{0.4in}

\begin{abstract}
The purely affine Lagrangian for linear electrodynamics, that has the form of the Maxwell Lagrangian in which the metric tensor is replaced by the symmetrized Ricci tensor and the electromagnetic field tensor by the tensor of homothetic curvature, is dynamically equivalent to the Einstein--Maxwell equations in the metric--affine and metric formulation.
We show that this equivalence is related to the invariance of the Maxwell Lagrangian under conformal transformations of the metric tensor.
We also apply to a purely affine Lagrangian the Legendre transformation with respect to the tensor of homothetic curvature to show that the corresponding Legendre term and the new Hamiltonian density are related to the Maxwell--Palatini Lagrangian for the electromagnetic field.
Therefore the purely affine picture, in addition to generating the gravitational Lagrangian that is linear in the curvature, justifies why the electromagnetic Lagrangian is quadratic in the electromagnetic field.
\end{abstract}


\maketitle

\section{Introduction}
\label{secIntro}

In the {\em purely affine} (Einstein--Eddington) formulation of general relativity\cite{Ein,Edd1,Edd2,Schr2,Kij}, a Lagrangian density depends on a torsionless affine connection and the symmetric part of the Ricci tensor.
This formulation constructs the symmetric metric tensor from the derivative of the Lagrangian density with respect to the symmetrized Ricci tensor, obtaining an algebraic relation between these two tensors.
It derives the field equations by varying the total action with respect to the connection, which gives a differential relation between the connection and the metric tensor.
This relation yields a differential equation for the metric.
In the {\em metric--affine} (Einstein--Palatini) formulation\cite{Schr2,Pal}, both the metric tensor and the torsionless connection are independent variables, and the field equations are derived by varying the action with respect to these quantities.
The corresponding Lagrangian density is linear in the symmetric part of the Ricci tensor.
In the {\em purely metric} (Einstein--Hilbert) formulation\cite{Hilb1,Hilb2,LL2}, the metric tensor is a variable, the affine connection is the Levi-Civita connection of the metric and the field equations are derived by varying the action with respect to the metric tensor.
The corresponding Lagrangian density is linear in the Ricci scalar.

Ferraris and Kijowski showed that all three formulations of general relativity are dynamically equivalent and the relation between them is analogous to the Legendre relation between Lagrangian and Hamiltonian dynamics\cite{FK2}.
This statement can be generalized to theories of gravitation with purely affine Lagrangians that depend on the full Ricci tensor and the tensor of homothetic curvature\cite{univ,nonsym}, and to a general connection with torsion\cite{nonsym}.
In the purely affine formulation of gravity, the connection is the fundamental variable, analogous to the coordinates in relativistic mechanics.
Consequently, the curvature corresponds to the four-dimensional velocities, and the metric corresponds to the generalized momenta\cite{Kij}.
The relation between the purely affine and metric--affine picture of general relativity shows that the metric--affine Lagrangian density for the gravitational field is a Legendre term corresponding to the scalar product of the velocities and momenta in classical mechanics\cite{Kij}.
The metric--affine and purely metric Lagrangian density for the gravitational field automatically turn out to be linear in the curvature tensor\cite{FK3}.

Ferraris and Kijowski also found that the purely affine Lagrangian for the electromagnetic field, that has the form of the Maxwell Lagrangian in which the metric tensor is replaced by the symmetrized Ricci tensor, is dynamically equivalent to the Einstein--Maxwell Lagrangian in the metric--affine and metric formulation\cite{FK1}.
This equivalence was proven by transforming to a reference system in which the electric and magnetic vectors (at the given point in spacetime) are parallel to one another.
Such a transformation is always possible except for the case when these vectors are mutually perpendicular and equal in magnitude\cite{LL2}.
This case, describing for example a plane electromagnetic wave, was examined in Ref.\cite{Niko2} and completed the proof.

In general relativity, the electromagnetic field and its sources are considered to be on the side of the matter tensor in the field equations, i.e. they act as sources of the gravitational field.
In unified field theory, the electromagnetic field obtains the same geometric status as the gravitational field\cite{Goe}.
A general affine connection has enough degrees of freedom to make it possible to describe the classical gravitational and electromagnetic fields and the purely affine formulation of gravity allows an elegant unification of these fields.
Ferraris and Kijowski showed that while the gravitational field in the purely affine formulation is represented by the symmetric part of the Ricci tensor for a connection that is not restricted to be metric compatible and symmetric, the electromagnetic field can be represented by the tensor of homothetic curvature\cite{FK2}.
Such a construction is dynamically equivalent to the sourceless Einstein--Maxwell equations\cite{FK2} and can be generalized to sources\cite{unif}.
Ponomarev and Obukhov used the same construction in the metric--affine formulation\cite{PO}.

Purely affine Lagrangians that depend explicitly on a general, unconstrained affine connection and the symmetric part of the Ricci tensor are subject to an unphysical constraint on the source density, as in the metric--affine formulation of gravity\cite{San}.
The inclusion of the tensor of homothetic curvature, which is related to the electromagnetic field, in a purely affine Lagrangian replaces this constraint with the Maxwell equations and preserves the projective invariance of the Lagrangian without constraining the connection\cite{unif}.

The structure of this paper is the following.
In Sec.~\ref{secField} we review the purely affine formulation of gravity and its relation to the metric--affine and metric picture.
In Sec.~\ref{secEM} we show that the equivalence between the Ferraris--Kijowski and Maxwell Lagrangian results from the invariance of the latter under conformal transformations of the metric tensor.
In Sec.~\ref{secLeg} we adopt the unified framework for the classical gravitational and electromagnetic fields\cite{FK2,unif}, and apply the Legendre transformation with respect to the tensor of homothetic curvature.
We show that the corresponding Legendre term and the new Hamiltonian density are directly related to the two terms in the Maxwell--Palatini Lagrangian for the electromagnetic field\cite{ADM}.
The simplest form of this Hamiltonian density yields the linear Maxwell electrodynamics.
Therefore the purely affine picture not only naturally derives the gravitational Lagrangian that is linear in the curvature, but also justifies why the electromagnetic Lagrangian is quadratic in the electromagnetic field.
We briefly discuss and summarize the results in Sec.~\ref{secSum}.

\section{Field Equations}
\label{secField}

A general purely affine Lagrangian density $\textgoth{L}$ depends on the affine connection $\Gamma^{\,\,\rho}_{\mu\,\nu}$ and the curvature tensor, $R^\rho_{\phantom{\rho}\mu\sigma\nu}=\Gamma^{\,\,\rho}_{\mu\,\nu,\sigma}-\Gamma^{\,\,\rho}_{\mu\,\sigma,\nu}+\Gamma^{\,\,\kappa}_{\mu\,\nu}\Gamma^{\,\,\rho}_{\kappa\,\sigma}-\Gamma^{\,\,\kappa}_{\mu\,\sigma}\Gamma^{\,\,\rho}_{\kappa\,\nu}$.\footnote{
A purely affine Lagrangian can also depend on derivatives of the curvature tensor.
We restrict attention to Lagrangians that depend only on $\Gamma^{\,\,\rho}_{\mu\,\nu}$ and $R^\rho_{\phantom{\rho}\mu\sigma\nu}$ since the corresponding field equations are second order.
}
We assume that the dependence of the Lagrangian on the curvature is restricted to the contracted curvature tensors\cite{nonsym}, of which there exist three: the symmetric $P_{\mu\nu}=R_{(\mu\nu)}$ and antisymmetric $R_{[\mu\nu]}$ part of the Ricci tensor, $R_{\mu\nu}=R^\rho_{\phantom{\rho}\mu\rho\nu}$, and the antisymmetric tensor of {\em homothetic curvature} (second Ricci tensor), $Q_{\mu\nu}=R^\rho_{\phantom{\rho}\rho\mu\nu}=\Gamma^{\,\,\rho}_{\rho\,\nu,\mu}-\Gamma^{\,\,\rho}_{\rho\,\mu,\nu}$, which has the form of a curl\cite{Schr2,Scho}.
The metric structure associated with a purely affine Lagrangian is obtained using\cite{Edd1,Edd2,Kij,FK2,FK3,FK1,Niko1}:
\begin{equation}
{\sf g}^{\mu\nu}\equiv-2\kappa\frac{\partial\textgoth{L}}{\partial P_{\mu\nu}},
\label{met1}
\end{equation}
where ${\sf g}^{\mu\nu}$ is the symmetric fundamental tensor density and $\kappa=\frac{8\pi G}{c^4}$.
The symmetric contravariant metric tensor is defined by
\begin{equation}
g^{\mu\nu}\equiv\frac{{\sf g}^{\mu\nu}}{\sqrt{-\mbox{det}{\sf g}^{\rho\sigma}}}.
\label{met2}
\end{equation}
For purely affine Lagrangians that do not depend on $R_{[\mu\nu]}$ these definitions are equivalent to those in Schr\"{o}dinger's nonsymmetric purely affine gravity\cite{Schr2,Schr1,Niko0}: ${\sf g}^{\mu\nu}=-2\kappa\frac{\partial\textgoth{L}}{\partial R_{\mu\nu}}$ and $g^{\mu\nu}=\frac{{\sf g}^{(\mu\nu)}}{\sqrt{-\mbox{det}{\sf g}^{(\rho\sigma)}}}$.
To make this definition meaningful, we have to assume $\mbox{det}({\sf g}^{(\mu\nu)})\neq0$, which also guarantees that the tensor $g^{\mu\nu}$ has the Lorentzian signature $(+,-,-,-)$ everywhere if its signature is Lorentzian at one point\cite{FK3}.
The covariant metric tensor $g_{\mu\nu}$ is related to the contravariant metric tensor by $g^{\mu\rho}g_{\nu\rho}=\delta^\mu_\nu$.
The tensors $g^{\mu\nu}$ and $g_{\mu\nu}$ are used for raising and lowering indices.
We also define an antisymmetric tensor density:
\begin{equation}
{\sf h}^{\mu\nu}\equiv-2\kappa\frac{\partial\textgoth{L}}{\partial Q_{\mu\nu}},
\label{amet1}
\end{equation}
and the {\em hypermomentum} density conjugate to the affine connection\cite{HK,HLS}:
\begin{equation}
\Pi_{\phantom{\mu}\rho\phantom{\nu}}^{\mu\phantom{\rho}\nu}\equiv-2\kappa\frac{\partial\textgoth{L}}{\partial \Gamma^{\,\,\rho}_{\mu\,\nu}},
\label{con1}
\end{equation}
which has the same dimension as the connection.

If we do not restrict the connection $\Gamma^{\,\,\rho}_{\mu\,\nu}$ to be symmetric, the variation of the Ricci tensor can be transformed into the variation of the connection by means of the Palatini formula\cite{Schr2}: $\delta R_{\mu\nu}=\delta\Gamma^{\,\,\rho}_{\mu\,\nu;\rho}-\delta\Gamma^{\,\,\rho}_{\mu\,\rho;\nu}-2S^\sigma_{\phantom{\sigma}\rho\nu}\delta\Gamma^{\,\,\rho}_{\mu\,\sigma}$, where $S^\rho_{\phantom{\rho}\mu\nu}=\Gamma^{\,\,\,\,\rho}_{[\mu\,\nu]}$ is the torsion tensor and the semicolon denotes the covariant differentiation with respect to $\Gamma^{\,\,\rho}_{\mu\,\nu}$.
Let us assume that the Lagrangian density $\textgoth{L}$ does not depend on $R_{[\mu\nu]}$.
The principle of least action $\delta S=0$, where $S=\frac{1}{c}\int d^4x\textgoth{L}(\Gamma^{\,\,\rho}_{\mu\,\nu},P_{\mu\nu},Q_{\mu\nu})$, with the variation with respect to $\Gamma^{\,\,\rho}_{\mu\,\nu}$ yield\cite{unif}:
\begin{equation}
{\sf g}^{\mu\nu}_{\phantom{\mu\nu},\rho}+\,^\ast\Gamma^{\,\,\mu}_{\sigma\,\rho}{\sf g}^{\sigma\nu}+\,^\ast\Gamma^{\,\,\nu}_{\rho\,\sigma}{\sf g}^{\mu\sigma}-\,^\ast\Gamma^{\,\,\sigma}_{\sigma\,\rho}{\sf g}^{\mu\nu}=\Pi_{\phantom{\mu}\rho\phantom{\nu}}^{\mu\phantom{\rho}\nu}-\frac{1}{3}\Pi_{\phantom{\mu}\sigma\phantom{\sigma}}^{\mu\phantom{\sigma}\sigma}\delta^\nu_\rho+2{\sf h}^{\nu\sigma}_{\phantom{\nu\sigma},\sigma}\delta^\mu_\rho-\frac{2}{3}{\sf h}^{\mu\sigma}_{\phantom{\mu\sigma},\sigma}\delta^\nu_\rho,
\label{field2}
\end{equation}
where $^\ast\Gamma^{\,\,\rho}_{\mu\,\nu}=\Gamma^{\,\,\rho}_{\mu\,\nu}+\frac{2}{3}\delta^\rho_\mu S_\nu$\cite{Schr2,Schr1} and $S_\mu=S^\nu_{\phantom{\nu}\mu\nu}$ is the torsion vector.
Antisymmetrizing and contracting the indices $\mu$ and $\rho$ in Eq.~(\ref{field2}) gives
\begin{equation}
{\sf h}^{\sigma\nu}_{\phantom{\sigma\nu},\sigma}={\sf j}^\nu,
\label{Max1}
\end{equation}
where
\begin{equation}
{\sf j}^\nu\equiv\frac{1}{8}\Pi_{\phantom{\sigma}\sigma\phantom{\nu}}^{\sigma\phantom{\sigma}\nu}.
\label{Max2}
\end{equation}
Eq.~(\ref{Max1}) has the form of the Maxwell equations for the electromagnetic field\cite{unif}.

The hypermomentum density $\Pi_{\phantom{\mu}\rho\phantom{\nu}}^{\mu\phantom{\rho}\nu}$ represents the {\em source} for the purely affine field equations\cite{unif}.
Since the tensor density ${\sf h}^{\mu\nu}$ is antisymmetric, the current vector density ${\sf j}^\mu$ must be conserved: ${\sf j}^\mu_{\phantom{\mu},\mu}=0$, which constrains how the connection $\Gamma^{\,\,\rho}_{\mu\,\nu}$ can enter a purely affine Lagrangian density $\textgoth{L}$: $\Pi_{\phantom{\sigma}\sigma\phantom{\nu},\nu}^{\sigma\phantom{\sigma}\nu}=0$.
This conservation is valid even if $\textgoth{L}$ depends on $R_{[\mu\nu]}$ since ${\sf h}^{\mu\nu}$ would only be replaced by ${\sf h}^{\mu\nu}+{\sf k}^{\mu\nu}$, where ${\sf k}^{\mu\nu}\equiv-2\kappa\frac{\partial\textgoth{L}}{\partial R_{[\mu\nu]}}$.
If $\textgoth{L}$ depends neither on $Q_{\mu\nu}$ or $R_{[\mu\nu]}$, the field equation~(\ref{Max1}) becomes a stronger, algebraic constraint on how the Lagrangian depends on the connection: $\Pi_{\phantom{\sigma}\sigma\phantom{\nu}}^{\sigma\phantom{\sigma}\nu}=0$.
The dependence of a purely affine Lagrangian on the tensor of homothetic curvature and the antisymmetric part of the Ricci tensor replaces this unphysical constraint with a field equation for ${\sf h}^{\mu\nu}$ or ${\sf g}^{[\mu\nu]}$, respectively.
Thus physical Lagrangians that depend explicitly on the affine connection should also depend on either $Q_{\mu\nu}$ or $R_{[\mu\nu]}$, which preserves the projective invariance of the total action without constraining the connection\cite{unif}.

If we apply to $\textgoth{L}(\Gamma^{\,\,\rho}_{\mu\,\nu},P_{\mu\nu},Q_{\mu\nu})$ the Legendre transformation with respect to $P_{\mu\nu}$\cite{Kij,FK3}, defining the Hamiltonian density $\textgoth{H}$:
\begin{equation}
\textgoth{H}\equiv\textgoth{L}-\frac{\partial\textgoth{L}}{\partial P_{\mu\nu}}P_{\mu\nu}=\textgoth{L}+\frac{1}{2\kappa}{\sf g}^{\mu\nu}P_{\mu\nu},
\label{Leg1}
\end{equation}
we find that $\textgoth{H}$ is a function of $\Gamma^{\,\,\rho}_{\mu\,\nu}$, ${\sf g}^{\mu\nu}$ and $Q_{\mu\nu}$.
The action becomes
\begin{equation}
S=\frac{1}{c}\int d^4x\Bigl(-\frac{1}{2\kappa}{\sf g}^{\mu\nu}P_{\mu\nu}+\textgoth{H}(\Gamma^{\,\,\rho}_{\mu\,\nu},{\sf g}^{\mu\nu},Q_{\mu\nu})\Bigr).
\label{act2}
\end{equation}
The action variation with respect to ${\sf g}^{\mu\nu}$ yields the first Hamilton equation\cite{Kij,FK3,unif}:
\begin{equation}
P_{\mu\nu}=2\kappa\frac{\partial\textgoth{H}}{\partial {\sf g}^{\mu\nu}}.
\label{Ham1}
\end{equation}
The variations with respect to $P_{\mu\nu}$ and $Q_{\mu\nu}$ can be transformed to the variation with respect to $\Gamma^{\,\,\rho}_{\mu\,\nu}$ by means of the Palatini formula and the variation of a curl, respectively, giving the second Hamilton equation equivalent to the field equations~(\ref{field2}).

The analogous transformation in classical mechanics goes from a {\em Lagrangian} $L(q^i,\dot{q}^i)$ to a {\em Hamiltonian} $H(q^i,p^i)=p^j\dot{q}^j-L(q^i,\dot{q}^i)$ (or, more precisely, a Routhian since not all the variables are subject to a Legendre transformation\cite{LL1}) with $p^i=\frac{\partial{L}}{\partial\dot{q}^i}$, where the tensor $P_{\mu\nu}$ corresponds to generalized velocities $\dot{q}^i$ and the density ${\sf g}^{\mu\nu}$ to canonical momenta $p^i$\cite{Kij,FK3}.
Accordingly, the affine connection plays the role of the configuration $q^i$ and the source density $\Pi_{\phantom{\mu}\rho\phantom{\nu}}^{\mu\phantom{\rho}\nu}$ corresponds to generalized forces $f^i=\frac{\partial{L}}{\partial q^i}$\cite{Kij}.
The field equations~(\ref{field2}) correspond to the Lagrange equations which result from Hamilton's principle $\delta\int L(q^i,\dot{q}^i)dt=0$ for arbitrary variations $\delta q^i$ vanishing at the boundaries of the configuration, while the Hamilton equations result from the same principle written as $\delta\int(p^j\dot{q}^j-H(q^i,p^i))dt=0$ for arbitrary variations $\delta q^i$ and $\delta p^i$\cite{LL1}.
The field equations~(\ref{field2}) correspond to the second Hamilton equation, $\dot{p}^i=-\frac{\partial H}{\partial q^i}$, and Eq.~(\ref{Ham1}) to the first Hamilton equation, $\dot{q}^i=\frac{\partial H}{\partial p^i}$.

If we identify $\textgoth{H}$ with the Lagrangian density for matter $\mathcal{L}_{MA}$ in the metric--affine formulation of gravitation\cite{FK3}\footnote{
This identification means that the purely affine theory based on the Lagrangian $\textgoth{L}$ and the metric--affine theory based on the matter Lagrangian $\mathcal{L}_{MA}$ equal to $\textgoth{H}$ have the same solutions to the corresponding field equations.
}
then Eq.~(\ref{Ham1}) has the form of the Einstein equations of general relativity,
\begin{equation}
P_{\mu\nu}-\frac{1}{2}Pg_{\mu\nu}=\kappa T_{\mu\nu},
\label{Ein}
\end{equation}
where $P=P_{\mu\nu}g^{\mu\nu}$ and the symmetric energy--momentum tensor $T_{\mu\nu}$ is defined by the variational relation: $2\kappa\delta\mathcal{L}_{MA}=T_{\mu\nu}\delta{\sf g}^{\mu\nu}$.
From Eq.~(\ref{act2}) it follows that $-\frac{1}{2\kappa}P\sqrt{-g}$, where $g=\mbox{det}g_{\mu\nu}$, is the metric--affine Lagrangian density for the gravitational field $\mathcal{L}_g$, in agreement with the general-relativistic form.
The transition from the affine to the metric--affine formalism shows that the gravitational Lagrangian density $\mathcal{L}_g$ is a {\em Legendre term} corresponding to $p^j\dot{q}^j$ in classical mechanics\cite{Kij}.
Therefore the purely affine and metric--affine formulation of gravitation are dynamically equivalent if $\textgoth{L}$ depends on the affine connection, the symmetric part of the Ricci tensor\cite{FK3} and the tensor of homothetic curvature\cite{nonsym}.

Substituting Eq.~(\ref{Max2}) to Eq.~(\ref{field2}) gives a linear algebraic equation for $^\ast\Gamma^{\,\,\,\,\rho}_{\mu\,\nu}$ as a function of the metric tensor, its first derivatives and the density $\Pi_{\phantom{\mu}\rho\phantom{\nu}}^{\mu\phantom{\rho}\nu}$.
We decompose the connection $^\ast\Gamma^{\,\,\rho}_{\mu\,\nu}$ as
\begin{equation}
^\ast\Gamma^{\,\,\rho}_{\mu\,\nu}=\{^{\,\,\rho}_{\mu\,\nu}\}_g+V^\rho_{\phantom{\rho}\mu\nu},
\label{sol1}
\end{equation}
where $\{^{\,\,\rho}_{\mu\,\nu}\}_g$ is the Christoffel connection of the metric tensor $g_{\mu\nu}$ and $V^\rho_{\phantom{\rho}\mu\nu}$ is a tensor.
Consequently, the Ricci tensor of the affine connection $\Gamma^{\,\,\rho}_{\mu\,\nu}$ is given by\cite{Scho}
\begin{equation}
R_{\mu\nu}(\Gamma)=R_{\mu\nu}(g)-\frac{2}{3}(S_{\nu:\mu}-S_{\mu:\nu})+V^\rho_{\phantom{\rho}\mu\nu:\rho}-V^\rho_{\phantom{\rho}\mu\rho:\nu}+V^\sigma_{\phantom{\sigma}\mu\nu}V^\rho_{\phantom{\rho}\sigma\rho}-V^\sigma_{\phantom{\sigma}\mu\rho}V^\rho_{\phantom{\rho}\sigma\nu},
\label{sol2}
\end{equation}
where $R_{\mu\nu}(g)$ is the Riemannian Ricci tensor of the metric tensor $g_{\mu\nu}$ and the colon denotes the covariant differentiation with respect to $\{^{\,\,\rho}_{\mu\,\nu}\}_g$.
Eq.~(\ref{Ein}) and symmetrized Eq.~(\ref{sol2}) give
\begin{equation}
R_{\mu\nu}(g)=\kappa T_{\mu\nu}-\frac{\kappa}{2}T_{\rho\sigma}g^{\rho\sigma}g_{\mu\nu}-V^\rho_{\phantom{\rho}(\mu\nu):\rho}+V^\rho_{\phantom{\rho}(\mu|\rho|:\nu)}-V^\sigma_{\phantom{\sigma}(\mu\nu)}V^\rho_{\phantom{\rho}\sigma\rho}+V^\sigma_{\phantom{\sigma}(\mu|\rho}V^\rho_{\phantom{\rho}\sigma|\nu)}.
\label{EMT}
\end{equation}
We also have\cite{Scho}
\begin{equation}
Q_{\mu\nu}=-\frac{8}{3}(S_{\nu,\mu}-S_{\mu,\nu})+V^\rho_{\phantom{\rho}\rho\nu,\mu}-V^\rho_{\phantom{\rho}\rho\mu,\nu}.
\label{q1}
\end{equation}
The general solution for the tensor $V^\rho_{\phantom{\rho}\mu\nu}$ is given in Refs.\cite{unif,PO}.
If there are no sources, $\Pi_{\phantom{\mu}\rho\phantom{\nu}}^{\mu\phantom{\rho}\nu}=0$, the connection $\Gamma^{\,\,\rho}_{\mu\,\nu}$ depends only on the metric tensor $g_{\mu\nu}$ representing a free gravitational field and the torsion vector $S_\mu$ corresponding to the vectorial degree of freedom associated with the projective invariance\cite{unif}.

The purely metric formulation of gravitation is dynamically equivalent to the purely affine and metric--affine formulation, which can be shown by applying to $\textgoth{H}(\Gamma^{\,\,\rho}_{\mu\,\nu},{\sf g}^{\mu\nu},Q_{\mu\nu})$ the Legendre transformation with respect to $\Gamma^{\,\,\rho}_{\mu\,\nu}$\cite{FK3}.
This transformation defines the Lagrangian density in the momentum space $\textgoth{K}$:
\begin{equation}
\textgoth{K}\equiv\textgoth{H}-\frac{\partial\textgoth{H}}{\partial\Gamma^{\,\,\rho}_{\mu\,\nu}}\Gamma^{\,\,\rho}_{\mu\,\nu}=\textgoth{H}+\frac{1}{2\kappa}\Pi_{\phantom{\mu}\rho\phantom{\nu}}^{\mu\phantom{\rho}\nu}\Gamma^{\,\,\rho}_{\mu\,\nu},
\label{Leg2}
\end{equation}
which is a function of ${\sf g}^{\mu\nu}$, $\Pi_{\phantom{\mu}\rho\phantom{\nu}}^{\mu\phantom{\rho}\nu}$ and $Q_{\mu\nu}$.
The action variation with respect to ${\sf g}^{\mu\nu}$ yields the Einstein equations:
\begin{equation}
P_{\mu\nu}=2\kappa\frac{\partial\textgoth{K}}{\partial{\sf g}^{\mu\nu}}.
\label{Ham2}
\end{equation}
The variations with respect to $P_{\mu\nu}$ and $Q_{\mu\nu}$ can be transformed to the variation with respect to $\Gamma^{\,\,\rho}_{\mu\,\nu}$ by means of the Palatini formula and the variation of a curl, respectively, giving the field equations~(\ref{field2}).
Finally, the variation with respect to $\Pi_{\phantom{\mu}\rho\phantom{\nu}}^{\mu\phantom{\rho}\nu}$ gives
\begin{equation}
\Gamma^{\,\,\rho}_{\mu\,\nu}=2\kappa\frac{\partial\textgoth{K}}{\partial\Pi_{\phantom{\mu}\rho\phantom{\nu}}^{\mu\phantom{\rho}\nu}},
\label{Ham3}
\end{equation}
in accordance with Eq.~(\ref{Leg2}).

The analogous transformation in classical mechanics goes from a Hamiltonian $H(q^i,p^i)$ to a momentum Lagrangian $K(p^i,\dot{p}^i)=-f^j q^j-H(q^i,p^i)$.
The equations of motion result from Hamilton's principle written as $\delta\int(p^j\dot{q}^j+f^j q^j+K(p^i,\dot{p}^i))dt=0$.
The quantity $K$ is a Lagrangian with respect to $p^i$ because $p^j\dot{q}^j+f^j q^j$ is a total time derivative and does not affect the action variation.

If we define:
\begin{equation}
C_\rho^{\phantom{\rho}\mu\nu}\equiv\Pi_{\phantom{(\mu}\rho\phantom{\nu)}}^{(\mu\phantom{\rho}\nu)}-\frac{1}{3}\delta^{(\mu}_\rho\Pi_{\phantom{\nu)}\sigma\phantom{\sigma}}^{\nu)\phantom{\sigma}\sigma}-\frac{1}{6}\Pi_{\phantom{\sigma}\sigma\phantom{(\mu}}^{\sigma\phantom{\sigma}(\mu}\delta^{\nu)}_\rho-\,^\ast\Gamma^{\,\,\,\,\mu}_{(\sigma\,\rho)}{\sf g}^{\sigma\nu}-\,^\ast\Gamma^{\,\,\,\,\nu}_{(\rho\,\sigma)}{\sf g}^{\mu\sigma}+\,^\ast\Gamma^{\,\,\,\,\sigma}_{(\sigma\,\rho)}{\sf g}^{\mu\nu},
\end{equation}
where the connection $\Gamma^{\,\,\rho}_{\mu\,\nu}$ depends on the source density $\Pi_{\phantom{\mu}\rho\phantom{\nu}}^{\mu\phantom{\rho}\nu}$ via Eq.~(\ref{con1}) or Eq.~(\ref{Ham3}), then the field equations~(\ref{field2}) and~(\ref{Max2}) can be written as ${\sf g}^{\mu\nu}_{\phantom{\mu\nu},\rho}=C_\rho^{\phantom{\rho}\mu\nu}$.
Accordingly, $\Pi_{\phantom{\mu}\rho\phantom{\nu}}^{\mu\phantom{\rho}\nu}$ can be expressed in terms of ${\sf g}^{\mu\nu}_{\phantom{\mu\nu},\rho}$ and $S^\rho_{\phantom{\rho}\mu\nu}$.
Consequently, $\textgoth{K}(g_{\mu\nu},g_{\mu\nu,\rho},S^\rho_{\phantom{\rho}\mu\nu},Q_{\mu\nu})$ can be identified with a Lagrangian density for matter $\mathcal{L}_M$ in the purely metric formulation of general relativity with torsion.\footnote{
This identification means that the purely affine theory based on the Lagrangian $\textgoth{L}$ and the purely metric theory based on the matter Lagrangian $\mathcal{L}_M$ equal to $\textgoth{K}$ have the same solutions to the corresponding field equations.
}
Similarly, the tensor $P_{\mu\nu}(\Gamma^{\,\,\rho}_{\mu\,\nu})$ in Eq.~(\ref{Ham2}) can be expressed as $P_{\mu\nu}(g_{\mu\nu},g_{\mu\nu,\rho},S^\rho_{\phantom{\rho}\mu\nu})$ which can be decomposed into the Riemannian Ricci tensor $\mathcal{R}_{\mu\nu}$ and terms with the torsion tensor\cite{Scho}, yielding the standard form of the Einstein equations\cite{nonsym,FK3,unif}.
The equivalence of purely affine gravity with standard general relativity, which is a metric theory, implies that the former is consistent with experimental tests of the weak equivalence principle\cite{Wi}.

The metric--affine Lagrangian density for the gravitational field $\mathcal{L}_g$ automatically turns out to be {\em linear} in the curvature tensor.
The purely metric Lagrangian density for the gravitational field also turns out to be linear in the curvature tensor: ${\sf L}_g=-\frac{1}{2\kappa}\mathcal{R}\sqrt{-g}$, since $P$ is a linear function of $\mathcal{R}=\mathcal{R}_{\mu\nu}g^{\mu\nu}$.
Thus metric--affine and metric Lagrangians for the gravitational field that are nonlinear with respect to curvature cannot be derived from a purely affine Lagrangian that depends on the connection and the contracted curvature tensors.

\section{Electromagnetic Field}
\label{secEM}

The purely affine Lagrangian density of Ferraris and Kijowski\cite{FK1} is given by
\begin{equation}
\textgoth{L}_{EM}=-\frac{1}{4}\sqrt{-\wp}F_{\alpha\beta}F_{\rho\sigma}P^{\alpha\rho}P^{\beta\sigma},
\label{Lagr3}
\end{equation}
where $F_{\mu\nu}=A_{\nu,\mu}-A_{\mu,\nu}$ is the electromagnetic field tensor, $\wp=\mbox{det}P_{\mu\nu}$ and the tensor $P^{\mu\nu}$ is reciprocal to $P_{\mu\nu}$.
This Lagrangian density has the form of the metric\footnote{
Or metric--affine, since the Lagrangian~(\ref{Lagr3}) does not depend explicitly on the connection and thus both formulations are identical.
}
Maxwell Lagrangian density of the electromagnetic field:
\begin{equation}
\textgoth{H}_{EM}=-\frac{1}{4}\sqrt{-g}F_{\alpha\beta}F_{\rho\sigma}g^{\alpha\rho}g^{\beta\sigma},
\label{Lagr4}
\end{equation}
in which the covariant metric tensor is replaced by the symmetrized Ricci tensor $P_{\mu\nu}$.
Accordingly, the contravariant metric tensor is replaced by $P^{\mu\nu}$.
From Eq.~(\ref{Lagr4}) it follows that
\begin{equation}
P_{\mu\nu}-\frac{1}{2}Pg_{\mu\nu}=\kappa\biggl(\frac{1}{4}F_{\alpha\beta}F_{\rho\sigma}g^{\alpha\rho}g^{\beta\sigma}g_{\mu\nu}-F_{\mu\alpha}F_{\nu\beta}g^{\alpha\beta}\biggr),
\label{EinMax1}
\end{equation}
which yields $P=0$ and, due to Eq.~(\ref{Leg1}), $\textgoth{H}=\textgoth{L}$.
The Lagrangian~(\ref{Lagr3}) is dynamically equivalent to the Lagrangian~(\ref{Lagr4}), i.e. $\textgoth{H}=\textgoth{H}_{EM}$ is equivalent to $\textgoth{L}=\textgoth{L}_{EM}$.
This can be demonstrated by showing that the identity
\begin{equation}
\sqrt{-\wp}F_{\alpha\beta}P^{\alpha\rho}P^{\beta\sigma}=\sqrt{-g}F_{\alpha\beta}g^{\alpha\rho}g^{\beta\sigma}
\label{REM2}
\end{equation}
is valid in the frame of reference in which the electric and magnetic vectors (at the given point in spacetime) are parallel to one another\cite{FK1}, and for the special case when they are mutually perpendicular and equal in magnitude\cite{Niko2}.\footnote{
If we would like to show this equivalence generally, without transforming to a particular reference system in which the calculations are simpler, we would have to solve an algebraic equation for $P_{\mu\nu}$:
\begin{equation}
\sqrt{-g}g^{\mu\nu}=\kappa\sqrt{-\wp}\biggl(\frac{1}{4}P^{\mu\nu}P^{\beta\sigma}-P^{\mu\beta}P^{\nu\sigma}\biggr)P^{\alpha\rho}F_{\alpha\beta}F_{\rho\sigma},
\label{det4}
\end{equation}
which results from Eqs.~(\ref{met1}) and~(\ref{met2}), and encounter the difficulty of taking the determinant of the right-hand side of this equation.}
The identity~(\ref{REM2}) does not hold if we add to the expression~(\ref{Lagr3}) a term that depends on the Ricci tensor, e.g., the Eddington affine Lagrangian density for the cosmological constant, $\textgoth{L}_{\Lambda}=\frac{1}{\kappa\Lambda}\sqrt{-\wp}$\cite{Niko2}.

The equivalence of the Lagrangians~(\ref{Lagr3}) and~(\ref{Lagr4}) is related to the invariance of the latter under conformal transformations of the metric tensor,\footnote{
The invariance of the Lagrangian~(\ref{Lagr4}) under a conformal transformation~(\ref{conf1}) is related to the fact that the trace $T$ of the dynamical energy-momentum tensor of the electromagnetic field vanishes.
Since the energy-momentum tensor $T_{\mu\nu}$ is generated from the matter action $S_m$ by the metric tensor:
\begin{equation}
\delta S_m=\frac{1}{2c}\int d^4 x\sqrt{-g}\,T_{\mu\nu}\delta g^{\mu\nu},
\label{EMT1}
\end{equation}
the action is invariant under an infinitesimal conformal transformation $\delta g^{\mu\nu}=\tilde{g}^{\mu\nu}-g^{\mu\nu}=(\Theta^{-1}-1)g^{\mu\nu}$ only if $T=T_{\mu\nu}g^{\mu\nu}=0$.}
\begin{equation}
g_{\mu\nu}\rightarrow\tilde{g}_{\mu\nu}=\Theta(x^\rho)g_{\mu\nu}.
\label{conf1}
\end{equation}
The transformation~(\ref{conf1}) yields:
\begin{eqnarray}
& & g^{\mu\nu}\rightarrow\tilde{g}^{\mu\nu}=\Theta^{-1}g^{\mu\nu}, \label{conf2a} \\
& & \sqrt{-g}\rightarrow\sqrt{-\tilde{g}}=\Theta^2\sqrt{-g}, \label{conf2b} \\
& & {\sf g}^{\mu\nu}\rightarrow\tilde{{\sf g}}^{\mu\nu}=\Theta{\sf g}^{\mu\nu}, \label{conf2c}
\end{eqnarray}
and the Lagrangian~(\ref{Lagr4}) is conformally invariant:
\begin{equation}
\textgoth{H}\rightarrow\tilde{\textgoth{H}}=\textgoth{H}.
\label{conf3}
\end{equation}
The identity $\delta\tilde{\textgoth{H}}=\delta\textgoth{H}$ can be written as
\begin{equation}
\tilde{P}_{\mu\nu}\delta\tilde{{\sf g}}^{\mu\nu}=P_{\mu\nu}\Theta^{-1}\delta\tilde{{\sf g}}^{\mu\nu}-P_{\mu\nu}{\sf g}^{\mu\nu}\Theta^{-1}\delta\Theta.
\label{conf4}
\end{equation}
For the Maxwell Lagrangian, the second term on the right-hand side of Eq.~(\ref{conf4}) vanishes, giving the transformation rule for the symmetrized Ricci tensor:
\begin{equation}
P_{\mu\nu}\rightarrow\tilde{P}_{\mu\nu}=\Theta^{-1}P_{\mu\nu}.
\label{confR}
\end{equation}
Consequently, we find:
\begin{eqnarray}
& & P^{\mu\nu}\rightarrow\tilde{P}^{\mu\nu}=\Theta P^{\mu\nu}, \label{conf5a} \\
& & \sqrt{-\wp}\rightarrow\sqrt{-\tilde{\wp}}=\sqrt{-\mbox{det}\tilde{P}_{\mu\nu}}=\Theta^{-2}\sqrt{-\wp}, \label{conf5b}
\end{eqnarray}
and the tensor density
\begin{equation}
{\textgoth{T}}^{\alpha\rho\beta\sigma}=\sqrt{-\wp}P^{\alpha\rho}P^{\beta\sigma}
\label{conf6}
\end{equation}
is invariant under the transformation~(\ref{confR}).

We know that for any matter Lagrangian in the metric formulation there exists the corresponding Lagrangian in the affine formulation\cite{FK2}.
We also know that the affine Lagrangian density that corresponds to the metric Maxwell Lagrangian must be, in order to give the field equations of linear electrodynamics, a quadratic function of the electromagnetic field tensor $F_{\mu\nu}$.
Finally, the affine Lagrangian density for the electromagnetic field must be invariant under the transformation~(\ref{confR}).
Since a quadratic function of the electromagnetic field tensor has four covariant indices, we need to contract it with a function of the symmetrized Ricci tensor that is a fourth-rank contravariant tensor density.
The tensor density~(\ref{conf6}) is the only function which satisfies this condition and is invariant under the transformation~(\ref{confR}).
Therefore, the purely affine Lagrangian density for the electromagnetic field is proportional to ${\textgoth{T}}^{\alpha\rho\beta\sigma}F_{\alpha\beta}F_{\rho\sigma}$, as in Eq.~(\ref{Lagr3}).
The factor $-1/4$ in Eq.~(\ref{Lagr3}) can be derived by the explicit calculation in a reference frame in which the electric and magnetic vectors are parallel\cite{FK1}.

\section{The Maxwell Lagrangian as Legendre Term}
\label{secLeg}

The purely affine formulation of gravity allows an elegant unification of the classical free electromagnetic and gravitational fields.
Ferraris and Kijowski showed that the gravitational field is represented by the symmetric part of the Ricci tensor for a connection that is not restricted to be metric compatible and symmetric, while the electromagnetic field can be represented by the tensor of homothetic curvature\cite{FK2}.
Such a construction is dynamically equivalent to the sourceless Einstein--Maxwell equations\cite{FK2} and can be generalized to sources\cite{unif}.

The formal similarity between $F_{\mu\nu}$ and the tensor of homothetic curvature $Q_{\mu\nu}$ (both tensors are curls) suggests that the purely affine Lagrangian density for the unified electromagnetic and gravitational fields is given by
\begin{equation}
\textgoth{L}_{EM}=-\frac{e^2}{4}\sqrt{-\wp}Q_{\alpha\beta}Q_{\rho\sigma}P^{\alpha\rho}P^{\beta\sigma},
\label{MaxA2}
\end{equation}
where $e$ has the dimension of electric charge\cite{FK2}.
Without loss of generality, $e$ can be taken equal to the charge of the electron.
The dynamical equivalence between this Lagrangian (found by Ferraris and Kijowski) and the Maxwell electrodynamics follows from the equivalence between the latter and $\textgoth{L}_{EM}$, since replacing $F_{\mu\nu}$ by $eQ_{\mu\nu}$ does not affect the algebraic relation of ${\sf g}^{\mu\nu}$ to $P_{\mu\nu}$ arising from Eqs.~(\ref{met1}) and~(\ref{Ham1}).
Eq.~(\ref{amet1}) gives
\begin{equation}
{\sf h}^{\mu\nu}=\kappa e^2\sqrt{-\wp}Q_{\alpha\beta}P^{\mu\alpha}P^{\nu\beta}.
\label{amet2}
\end{equation}
The torsion vector is related to the electromagnetic potential $A_\mu$ via Eq.~(\ref{q1}) and the correspondence relation $F_{\mu\nu}=eQ_{\mu\nu}$\cite{unif}:
\begin{equation}
S_\nu=\frac{3}{8}(-\frac{A_\nu}{e}+V^\rho_{\phantom{\rho}\rho\nu}).
\label{tor}
\end{equation}
If there are no sources, this vector is proportional to the electromagnetic potential.
This proportionality was assumed by Hammond in the Einstein--Maxwell formulation of gravitation and electromagnetism with propagating torsion\cite{Ham}.
Here, we assume the more general correspondence $Q_{\mu\nu}\propto F_{\mu\nu}$\cite{FK2} of which the relation $S_\mu\propto A_\mu$ is a special case.
The gauge transformation $A_\nu\rightarrow A_\nu+\partial_\nu \lambda$, where $\lambda$ is a scalar function of the coordinates, does not affect Eq.~(\ref{q1}).

Purely affine Lagrangians that depend explicitly on a general, unconstrained affine connection and the symmetric part of the Ricci tensor are subject to an unphysical constraint on the source density, as in the metric--affine formulation of gravity\cite{San}.
The inclusion of the tensor of homothetic curvature, which is related to the electromagnetic field, in a purely affine Lagrangian replaces this constraint with the Maxwell equations and preserves the projective invariance of the Lagrangian without constraining the connection\cite{unif}.

We now apply to the Hamiltonian density $\textgoth{H}(\Gamma^{\,\,\rho}_{\mu\,\nu},{\sf g}^{\mu\nu},Q_{\mu\nu})$ the Legendre transformation with respect to $Q_{\mu\nu}$.
This transformation defines a new density $\textgoth{F}$:
\begin{equation}
\textgoth{F}\equiv\textgoth{H}-\frac{\partial\textgoth{H}}{\partial Q_{\mu\nu}}Q_{\mu\nu}=\textgoth{H}+\frac{1}{2\kappa}{\sf h}^{\mu\nu}Q_{\mu\nu},
\label{Leg3}
\end{equation}
which is a function of $\Gamma^{\,\,\rho}_{\mu\,\nu}$, ${\sf g}^{\mu\nu}$ and ${\sf h}^{\mu\nu}$.
The action becomes
\begin{equation}
S=\frac{1}{c}\int d^4x\Bigl(-\frac{1}{2\kappa}P\sqrt{-g}-\frac{1}{2\kappa}{\sf h}^{\mu\nu}Q_{\mu\nu}+\textgoth{F}(\Gamma^{\,\,\rho}_{\mu\,\nu},{\sf g}^{\mu\nu},{\sf h}^{\mu\nu})\Bigr).
\label{act3}
\end{equation}
The action variation with respect to ${\sf g}^{\mu\nu}$ yields the Einstein equations:
\begin{equation}
P_{\mu\nu}=2\kappa\frac{\partial\textgoth{F}}{\partial{\sf g}^{\mu\nu}}.
\label{Ham4}
\end{equation}
The variation with respect to $\Gamma^{\,\,\rho}_{\mu\,\nu}$ gives the second pair of the Maxwell equations~(\ref{Max2}).
Finally, the variation with respect to ${\sf h}^{\mu\nu}$ gives
\begin{equation}
Q_{\mu\nu}=2\kappa\frac{\partial\textgoth{F}}{\partial{\sf h}^{\mu\nu}}.
\label{Ham5}
\end{equation}

The Maxwell--Palatini Lagrangian density for the electromagnetic field\cite{ADM} is given by
\begin{equation}
{\sf L}_{EM}=\sqrt{-g}\Bigl(A_{\nu,\mu}F^{\mu\nu}+\frac{1}{4}F_{\mu\nu}F^{\mu\nu}-A_\mu j^\mu\Bigr).
\label{MP1}
\end{equation}
The variation with respect to $A_\mu$ gives the second pair of the Maxwell equations $(\sqrt{-g}F^{\nu\mu})_{,\nu}=\sqrt{-g}j^\mu$, while the variation with respect to $F^{\mu\nu}$ yields the first pair $F_{\mu\nu}=A_{\nu,\mu}-A_{\mu,\nu}$.
Comparing Eqs.~(\ref{act3}) and~(\ref{MP1}) allows us to associate $eQ_{\mu\nu}$ with $A_{\nu,\mu}-A_{\mu,\nu}$ and $\frac{1}{\kappa e}{\sf h}^{\mu\nu}$ with $\sqrt{-g}F^{\mu\nu}$.\footnote{
Accordingly, we associate $e\Gamma^{\,\,\rho}_{\rho\,\nu}$ with $A_\nu$ (up to a gradient since $\Gamma^{\,\,\rho}_{\rho\,\nu}$ is not a tensor) and $\frac{1}{\kappa e}{\sf j}^\mu$ with $\sqrt{-g}j^\mu$.
}
Eq.~(\ref{Ham5}) reproduces the first pair of the Maxwell equations if $\textgoth{F}$ is a quadratic function of ${\sf h}^{\mu\nu}$ and does not depend on the affine connection:
\begin{equation}
\textgoth{F}=\frac{1}{4\kappa^2e^2\sqrt{-g}}{\sf h}^{\mu\nu}{\sf h}^{\alpha\beta}g_{\mu\alpha}g_{\nu\beta}.
\label{MP2}
\end{equation}
Comparing Eqs.~(\ref{con1}), (\ref{Max2}) and~(\ref{MP1}) indicates that the source density $\Pi_{\phantom{\mu}\rho\phantom{\nu}}^{\mu\phantom{\rho}\nu}$ must satisfy
\begin{equation}
\Pi_{\phantom{\mu}\rho\phantom{\nu}}^{\mu\phantom{\rho}\nu}=2\delta^\mu_\rho{\sf j}^\nu,
\label{cur}
\end{equation}
in agreement with the last footnote.
The Legendre term $\frac{1}{2\kappa}{\sf h}^{\mu\nu}Q_{\mu\nu}$ and the density $\textgoth{F}$ are directly related to the first two terms in the Maxwell--Palatini Lagrangian~(\ref{MP1}).

\section{Discussion and Summary}
\label{secSum}

The density $\textgoth{F}$ given by Eq.~(\ref{MP2}) is the simplest scalar density that can be constructed from the fields ${\sf h}^{\mu\nu}$ and $g_{\mu\nu}$.
The Legendre term $\frac{1}{2\kappa}{\sf h}^{\mu\nu}Q_{\mu\nu}$ and the density $\textgoth{F}$ are directly related to the two sourceless terms in the Maxwell--Palatini Lagrangian for the electromagnetic field\cite{ADM}.
Therefore the purely affine picture not only naturally derives the gravitational Lagrangian that is linear in the curvature and the second pair of the Maxwell equations, but also justifies why the electromagnetic Lagrangian is quadratic in the electromagnetic field that leads to the first pair of the Maxwell equations.\footnote{
The density~(\ref{MaxA2}), dynamically equivalent to~(\ref{MP2}), is the simplest scalar density that can be constructed from the tensors $Q_{\mu\nu}$ and $P_{\mu\nu}$.
}
If $\textgoth{F}$ is more complicated, so is the relation between the potential $A_\mu$ and the electromagnetic field tensor $F_{\mu\nu}$.
However, there always exists a function of $F_{\mu\nu}$ that is equal to the curl of $A_\mu$, in agreement with the electromagnetic gauge invariance.

A general affine connection has enough degrees of freedom to make it possible to unify the classical gravitational and electromagnetic fields in the purely affine formulation of gravity.
We reviewed the purely affine formulation of gravity and its relation to the metric--affine and metric picture, and showed that the equivalence between the Ferraris--Kijowski and Maxwell Lagrangian results from the invariance of the latter under conformal transformations of the metric tensor.
In the Ferraris--Kijowski unified model for the classical gravitational and electromagnetic fields\cite{FK2,unif} we applied to a purely affine Lagrangian the Legendre transformation with respect to the tensor of homothetic curvature.
We showed that the corresponding Legendre term and the new Hamiltonian density, taken to be of the simplest possible form, are directly related to the two sourceless terms in the Maxwell--Palatini Lagrangian for the linear Maxwell electrodynamics.
Therefore the purely affine picture not only explains why the gravitational Lagrangian is linear in the curvature, but also justifies why the electromagnetic Lagrangian is quadratic in the electromagnetic field.

The purely affine formulation of electromagnetism has one feature: in the zero-field limit, where $F_{\mu\nu}=0$, the Lagrangians~(\ref{Lagr3}) and~(\ref{MaxA2}) vanish, thus making impossible to apply Eq.~(\ref{met1}) to construct the metric tensor.
Therefore, there must exist a background field that depends on the tensor $P_{\mu\nu}$ so that the metric tensor is well-defined.
The simplest possibility, supported by recent astronomical observations, is the Eddington Lagrangian for the cosmological constant\cite{Edd1,Edd2,Schr2}.
The question of how to combine this Lagrangian with the Ferraris--Kijowski Lagrangian to obtain a viable model of gravitation and electromagnetism remains open\cite{Niko2}.

\section*{Acknowledgment}

The author would like to thank Ted Jacobson for valuable comments on the purely affine formulation of gravity.


\end{document}